Title: The Habitable Zone and Extreme Planetary Orbits

Authors: Stephen R. Kane, Dawn M. Gelino

Affiliation: NASA Exoplanet Science Institute, Caltech, MS 100-22, 770 South Wilson Avenue, Pasadena, CA 91125, USA

Correspondence: Stephen Kane, Address: NASA Exoplanet Science Institute, Caltech, MS 100-22, 770 South Wilson Avenue, Pasadena, CA 91125, USA, Phone: +1-626-395-1948, Fax: +1-626-397-7181, Email: skane@ipac.caltech.edu

Running title: Extreme Planetary Orbits

Abstract: The Habitable Zone for a given star describes the range of circumstellar distances from the star within which a planet could have liquid water on its surface, which depends upon the stellar properties. Here we describe the development of the Habitable Zone concept, its application to our own Solar System, and its subsequent application to exoplanetary systems. We further apply this to planets in extreme eccentric orbits and show how they may still retain life-bearing properties depending upon the percentage of the total orbit which is spent within the Habitable Zone.

# 1. Introduction

The detection of exoplanets has undergone extraordinary growth over the past couple of decades since sub-stellar companions were detected around HD 114762 (Latham et al 1989), pulsars (Wolszczan & Frail 1992), and 51 Peg (Mayor & Queloz 1995). From there the field rapidly expanded from the exclusivity of exoplanet detection to include exoplanet characterization. This has been largely facilitated by the discovery of transiting planets which allows unique opportunities to study the transmission and absorption properties of their atmospheres during primary transit (Agol et al 2010, Knutson et al 2007) and secondary eclipse (Charbonneau et al 2005, Richardson et al 2007). In addition, studies of phase curves allows insight into the thermal and albedo atmospheric properties (Kane & Gelino 2010, 2011). These studies are being applied to smaller planets as our detection sensitivity pushes down into the super-Earth regime (Bean et al 2011, Croll et al 2011).

The expansion of exoplanetary science is leading the field towards the question of the frequency of Earth-like planets and, in particular, of habitable planets. The merging of stellar and planetary properties results in a quantitative description of the Habitable Zone (HZ), usually defined as that region around a star where water can exist in a liquid state. Although habitability had certainly been considered before, the concept of a HZ first started to appear in the science literature during the 1950's in articles such those by Su-Shu Huang (1959, 1960). However, these were written at a time when atmospheric studies of other planets were in their infancy and the presence of planets outside of our Solar System had not yet been confirmed. The work of Kasting et al (1993) was instrumental in new advancements in HZ research and in particular quantifying the boundaries in terms of climate models as well as the properties of the host star.

Here we outline the development of the HZ and the factors which influence the boundary conditions. We apply this to our Solar System and discuss the extension to exoplanetary systems. We also discuss highly eccentric orbits and how the time spent within the HZ effects the conditions for habitability and extremophiles that could survive under these conditions.

## 2. Habitable Zone Boundaries

The boundaries of the HZ for a particular system are calculated based upon both the stellar properties and assumptions regarding the response of a planetary atmosphere to stellar flux. Calculations for HZ boundaries have undergone considerable change since the published articles of Huang (1959, 1960). Estimates for our Solar System by Dole & Asimov (1964) were in the relatively broad range of 0.725-1.24 AU, compared with the conservative estimates by Hart (1979) of 0.95-1.01 AU. Detailed models for runaway greenhouse implications for Venus were considered by Pollack (1971) and Kasting (1988). For more recent studies regarding the HZ, we refer the reader to the detailed models of Kasting et al

(1993), Kasting & Catling (2003), Underwood et al (2003), Selsis et al (2007), and Jones & Sleep (2010). Here we describe the primary factors which influence the boundaries and present examples both in and out of our Solar System.

The premise that remotely detectable habitability requires water to exist on a planetary surface in a liquid state allows a quantitative representation of the HZ to be estimated. This is established by combining the stellar properties with those of a hypothetical planet at various distances from the star. The flux received by the planet is a function of this distance and the luminosity of the star. Kasting et al (1993) used one-dimensional (altitude) climate models to calculate the atmospheric response of the atmosphere to the flux by considering conditions whereby the equilibrium would sway to a runaway greenhouse effect or to a runaway snowball effect. The runaway greenhouse cycle is summarized as: increased $CO_2$ and/or increased stellar flux; temperature increases; temperature rise leads to an increase in water vapor; increased greenhouse gas concentration leads to high temperatures. Note that the increase in water vapor leads to rapid loss of hydrogen and increased absorption at the wavelengths through which much of the Earth's Infra-Red radiation escapes, thus propagating the runaway warming of the surface. The runaway snowball model is summarized as: reduced greenhouse gases and/or reduced stellar flux; temperature decreases; resulting ice increases surface reflectivity; reduced heating leads to further ice formation. Note that this also results in significant amounts of atmospheric $CO_2$ condensing as dry ice and thus further reducing the greenhouse effect.

The overall habitability is further complicated by the various other factors inherent to the system. The flux from the host star depends on spectral type and luminosity class but also depends on the age of the star. As the age of the star increases, so does the luminosity and thus the HZ moves outwards from the center of the system. One then can consider a Continuously Habitable Zone (CHZ) in which a particular planetary orbit may remain while the star is on the main sequence, since the movement of the HZ continues even after the star leaves the main sequence (Lopez et al 2005). As described above, the planetary atmospheric and albedo properties play an important role in controlling the heat retained at the surface. The redistribution of that heat is in turn related to the planetary rotation rate, particularly important for planets in the HZ which may be tidally-locked around late-type stars (Williams & Kasting 1997). In addition, the planetary mass will influence such aspects as the amount of retained atmosphere and the level and continuity of volcanic activity. Too small a mass will likely lead to a lack of these attributes and too high a mass will lead to a possibly inhospitable thick hydrogen envelope that will further obscure the visibility of biosignatures. Thus, optimal planetary masses exist which result in a more suitable balancing of these effects.

In the following sections, we apply the equations of Underwood et al (2003) and Jones & Sleep (2010) which relate the HZ boundaries to stellar luminosity and effective temperature for reasonable models of planetary atmospheric responses.

## 2.1 Application to the Solar System

By way of demonstration, we first apply the HZ models to our own Solar System. For a solar effective temperature of 5778 K, we find the inner and outer boundaries of the HZ to be 0.836 AU and 1.656 AU respectively. Figure 1 shows the HZ depicted as a gray disk with the orbits of the terrestrial inner planets overlaid.

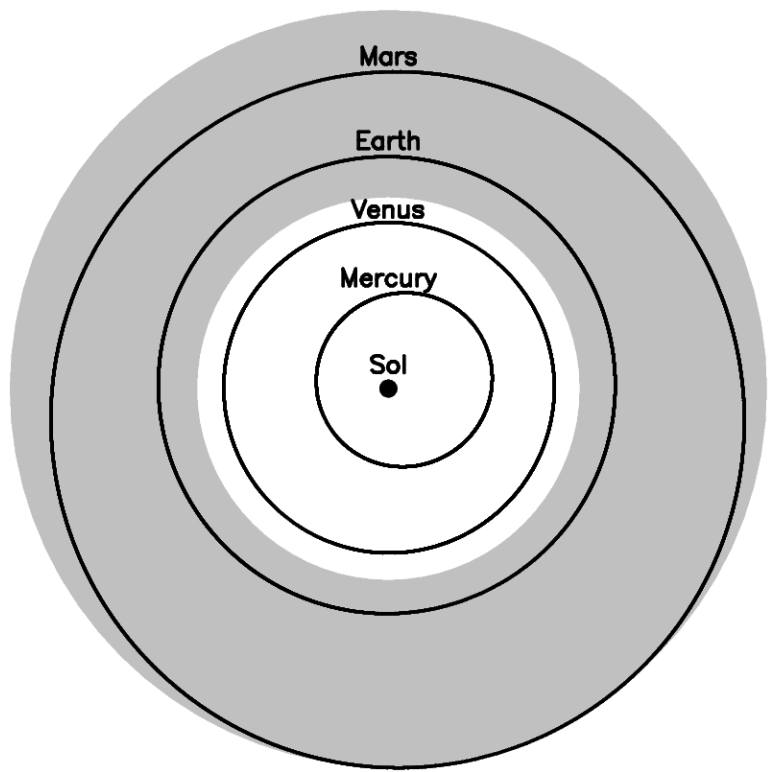

**Figure 1:** The inner Solar System showing the orbits of the terrestrial planets. The HZ is indicted by a gray disk.

An estimate of the equilibrium temperatures for the inner planets can be calculated by using their known orbital parameters and Bond albedos[1] and treating them as black bodies, as described by Kane & Gelino (2011). This calculation is further explored by considering two cases: one in which the heat is evenly redistributed throughout the surface and the other where the heat redistribution is restricted to the day side of the planet. The results of this are shown in Table 1, where Temp 1 is the former and Temp 2 is the latter. This shows that the model of complete heat redistribution works reasonably well for Solar System terrestrial planets. The major exception to this is Venus whose temperature has become dominated by the climate changes which have resulted in a runaway greenhouse effect, as described earlier. For Mercury, the actual surface temperature varies enormously is highly position dependent on the planet due to tidal locking and the lack of any substantial atmosphere. For Earth, the calculated temperature of 254 K compared with the actual temperature of 288 K reveals an inability of this simple model to correctly account for atmospheric (greenhouse) warming of the surface. Thus one should be cautious when using such a model to estimate temperatures of terrestrial-sized planets in exosystems since there can potentially be a large range of temperatures spanned by uncertainties in albedo and greenhouse effect.

---

1http://nssdc.gsfc.nasa.gov/planetary/factsheet/

| Planet | Temp 1 (K) | Temp 2 (K) | Actual Temp (K) |
|---|---|---|---|
| Mercury | 440 | 523 | 440 |
| Venus | 184 | 219 | 737 |
| Earth | 254 | 302 | 288 |
| Mars | 210 | 250 | 210 |

**Table 1:** Calculated black body temperatures for the terrestrial planets assuming complete heat redistribution (Temp 1) and no heat redistribution resulting in a hot day-side (Temp 2). For comparison, we show the actual measured mean surface temperatures.

## 2.2 Application to Exoplanets

These techniques may be applied to known exoplanetary systems for which there are a variety of host stars and associated planets. There have been numerous recent discoveries of super-Earth mass planets whose orbits have been found to lie within their stars' HZ. Examples include Gliese 581 d (von Paris et al 2010, Wordsworth et al 2011) which is in a multi-planet system around an M dwarf, and the more recent case of a potentially habitable planet around HD 85512 (Kaltenegger et al 2011). The investigation of the HZ for Kepler candidates is of particular interest since many of these planets fall in the terrestrial regime (Kaltenegger & Sasselov 2011).

Many of the known exoplanets are of Jovian mass and thus we do not necessarily consider them as habitable on their own. However, it is worth considering that these planets likely harbor their own systems of terrestrial moons considering the frequency with which we see such occurrences in our own system. Indeed one may consider how the ecology of the Jovian moons may have evolved, were Jupiter in the Earth's current orbit. The habitability of exomoons has been considered in the literature (Kaltenegger 2010, Porter & Grundy 2011) and attempts are being undertaken to detect their presence from Kepler mission data (Kipping et al 2009, 2012). There are also alternative energy sources for exomoons, such as tides, which may help or hinder habitability and even allow an extension of habitable conditions beyond the traditional HZ (Barnes et al 2009, Scharf 2006, Williams et al 1997).

**2.3 The Habitable Zone Gallery**

In order to provide an easily accessible service to the astronomical community for the investigation of HZ information for known exoplanetary systems, we developed the Habitable Zone Gallery (www.hzgallery.org). This service provides HZ details for each of the exoplanetary systems with sufficiently known planetary and stellar parameters to carry out the necessary calculations. This catalogue is regularly updated as new discoveries are made and new data becomes available. The service includes a sortable table with information on the percentage of the orbital phase each exoplanet spends within their HZ, planetary effective temperatures, and other basic planetary properties. In addition to the table, the service provides a gallery of known systems which plot the orbits and the location of the HZ with respect to those orbits, such as that seen in Figure 1, and movies which animate the movement and temperature of a planet as it orbits the star. A summary plot shows the relation between period, orbital eccentricity, and the percentage time spent in the HZ. The under-lying calculations and infrastructure for the Habitable Zone Gallery are described in detail by Kane & Gelino (2012). In the following section we consider extreme cases of eccentric orbits whose path takes the planet through the HZ.

## 3. Extreme Planetary Orbits

Here we describe cases where a planet moves in and out of the HZ due to the eccentricity of the orbit and the subsequent effects on habitability.

### 3.1 Frequency of Eccentric Planets

Amongst the surprises included in exoplanet discoveries is that of planets in highly eccentric orbits. Due to dynamical stability considerations, most of the very eccentric planets occupy single-planet systems. To show the frequency of these kinds of planetary orbits, we extracted data from the Exoplanet Orbit Database at exoplanets.org (Wright et al 2010). This resource includes only those planets which have complete orbital solutions which, at the time of writing, consists of 489 planets. Figure 2 plots the orbital eccentricity of these planets against the period. Of these, 187 (38%) have eccentricities larger than 0.2. There is a clear correlation with an increasing eccentricity dispersion with increasing period since short-period orbits have much smaller tidal circularization timescales (Goldreich & Soter 1966). Figure 3 shows the distributions of planetary masses over this same period range. One may thus be tempted to conclude from this figure that eccentric planets tend to be of Jovian-mass and at large semi-major axis. There is a selection effect however such that smaller mass planets at longer periods are more difficult to detect using radial velocity methods. Even so, these two figures do demonstrate that there are a large number of eccentric gas giants which fall within the expected period range for the HZ of main sequence stars.

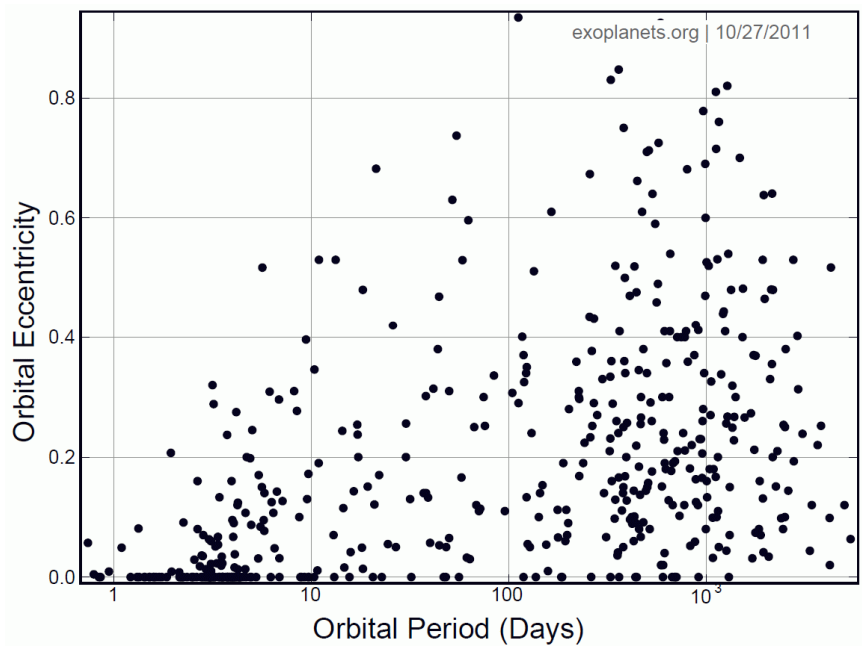

**Figure 2:** Orbital eccentricity vs period for known radial velocity planets.

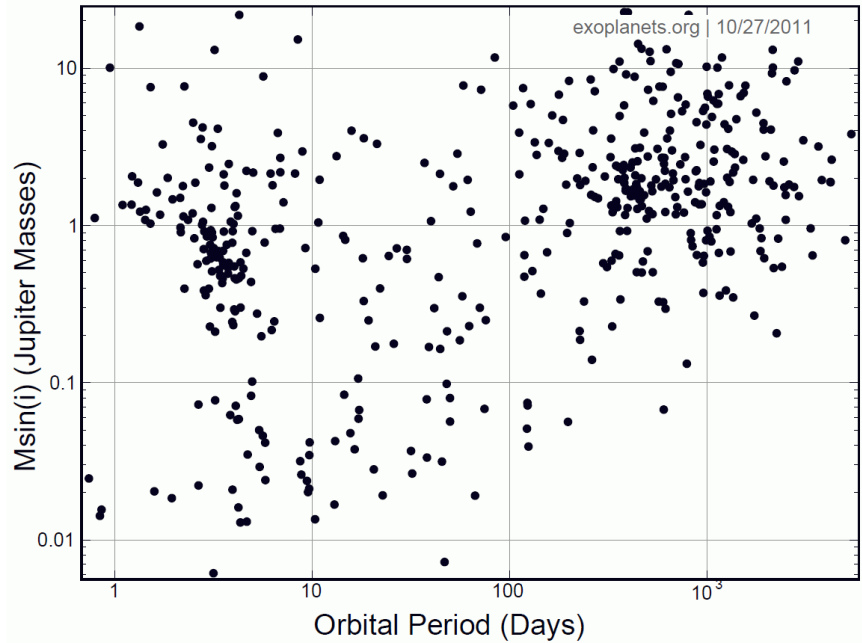

**Figure 3:** Planetary mass vs period for known radial velocity planets.

### 3.2 Orbital Passage Through the Habitable Zone

There have been several interesting studies regarding the habitability of terrestrial mass planets in eccentric orbits, such as that by Dressing (2010). Kita (2010) show the possible detrimental effects of terrestrial planet habitability when excited to an eccentric orbit by another companion in the system. Spiegel et al (2010) consider the effect of transiently eccentric orbits on planet habitability. In particular, the work of Williams & Pollard (2002) considers an orbit-averaged flux as a habitability diagnostic for eccentric orbits. We diverge from this metric since the average distance of the planet from the star is significantly farther than the difference between apastron and periastron and thus we use the percentage of time spent in the Habitable Zone where metabolic processes could proceed rather than relying on sustainable temperatures. However, the calculated equilibrium temperatures (as described in Section 2.1) make implicit assumptions regarding the response of the atmosphere to changes in flux. In reality, planetary temperatures at the surface and upper atmosphere are complicated functions of surface and atmospheric composition and dynamics as well as the long-term climate history and so can only be treated as a first-order approximation. See for example Williams & Kasting (1997), Spiegel et al (2998, 2009), Kane & Gelino (2011) and references therein. Here we briefly discuss two examples and we refer the reader to Kane & Gelino (2011, 2012) and the Habitable Zone Gallery for more detailed information.

One important consideration is the orbital stability of exomoons as the parent planet passes through periastron. Hamilton & Burns (1992) showed that the Hill radius at periastron is a good representation of the stability zone for satellites of planets in highly eccentric orbits. The giant planets discussed here will have Hill radii which extends beyond Galilean moon analogs and thus could retain similar moons under these conditions.

In Figures 4 and 5 we present two examples, one with periastron in the HZ and the other with apastron in the HZ (shown in gray). The planet orbiting HD 131664 is in a 1951 day orbit and has an eccentricity of 0.64. The minimum mass of the planet is 18 Jupiter masses resulting in a Hill Sphere radius of 0.2 AU (416 Jupiter radii) at periastron. The planet only spends 11% of the orbit in the HZ since it moves slowly near apastron. The calculated equilibrium temperature is 271 K at periastron and 127 K at apastron assuming a Bond albedo of 0 meaning that the planet and moons absorb 100% of the incident flux. Thus the planet and possible moons undergo long periods of hibernation conditions broken regularly with warmer habitable conditions (see Section 3.3). In fact the situation is improved for eccentric orbits through the extension of the outer HZ boundary as found by Williams & Pollard (2002) and Dressing et al (2010). In contrast, the planet orbiting HD 80606 is in a 111 day orbit with an eccentricity of 0.93 and is known to both transit and be eclipsed by its host star (Laughlin et al 2009). This planet has a mass of 3.9 Jupiter masses leading to a Hill Sphere radius of 0.0032 AU (6.7 Jupiter radii). In this case it is the apastron which lies within the HZ where it spends 40% of the orbital phase. The temperature at apastron is predicted as 286 K, however the close encounter with the star at periastron causes temperatures to reach a scorching 1546 K. This results in flash-heating of the upper atmosphere which doubles in temperature in only 6 hours as it passes through periastron. Thus organisms could only survive such environments if deeply buried under sufficient protective layers. For both of these examples, regular intervals of habitability for the moons depends strongly on the surface conditions and their response to the change in temperatures. For example, a water-rich moon will require sufficient time to melt out of a snowball state to allow metabolic processes to continue.

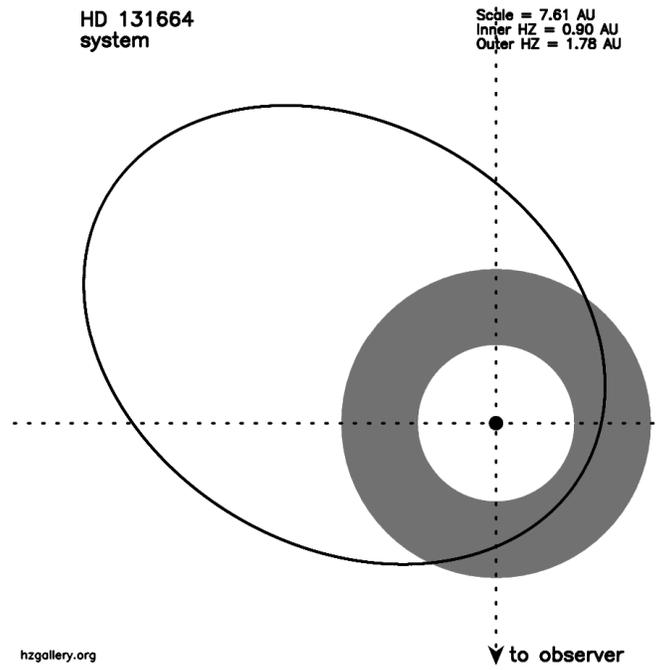

**Figure 4:** The planetary orbit and HZ for the HD 131664 system.

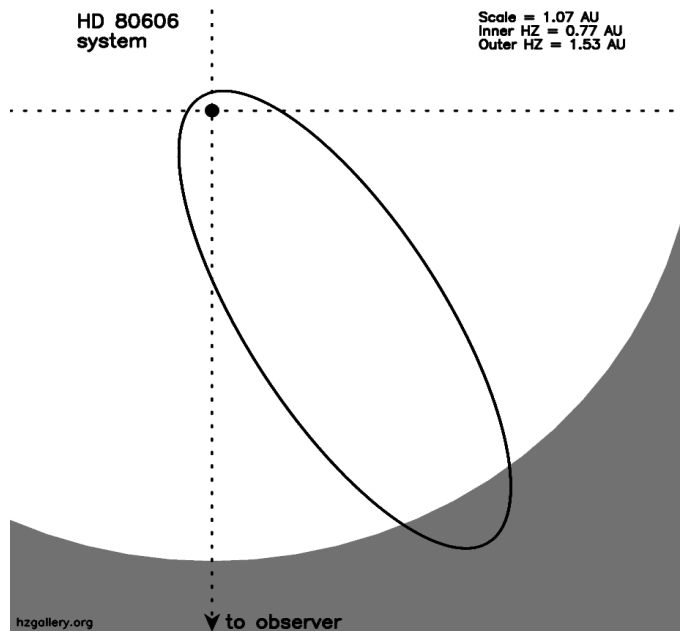

**Figure 5:** The planetary orbit and HZ for the HD 80606 system.

### 3.3 Hibernation and Sterilization

We now consider the survivability of extremophiles in these environments. A transition from liquid to frozen water and back again does not present an obstacle for most Earth-based microorganisms. It has been shown by de La Torre et al (2010) that organisms such as lichens and bacteria can survive for periods of at least 10 days when exposed to the harsh conditions outer space. Foucher et al (2010) presented the results of atmosphere re-entry experiments which demonstrate that microfossils can survive entry into Earth's atmosphere within sedimentary rock but that microorganisms require protection by at least 5 cm of rock for adequate shielding. During the time an exoplanet or exomoon spends exterior to the HZ, similar organisms can undergo metabolic slow-down for a period of hibernation without any immediate biological harm. The daily and annual biological clock we see exhibited by plants and cells on Earth could be adapted to different orbital durations to optimize time spent in favorable (habitable) conditions. Since we know that several terrestrial organisms can withstand substantial periods of time under extreme temperature conditions in combination with vacuum, UV-C irradiation and cosmic rays, then it is conceivable that such organisms could survive periastron flash-heating at a sufficient protective depth beneath the surface. In particular, oceans have a relatively high thermal inertia and could be maintained in liquid form even if the surface waters freeze over during the apastron winter. This can provide an environment where the annual range of temperatures is significantly less extreme than those experienced at the surface.

## 4. Conclusions

Although the concept of the HZ has been discussed for some time, it is only recently that sophisticated climate models are allowing concise quantification of this region. The study of exoplanetary atmospheres enables us to apply these concepts to known exosystems, even those with exoplanets in orbits which result in extreme temperature variations. A targeted search for exomoons in these environments may yield more surprises on what we consider "habitable".

## Acknowledgements

The authors would like to thank Lisa Kaltenegger, Ravikumar Kopparapu, and Rene Demets for several useful discussions. We would also like to thank the referees (Shawn Domagal-Goldman and two anonymous) for their feedback which greatly improved the manuscript. This research has made use of the Exoplanet Orbit Database and the Exoplanet Data Explorer at exoplanets.org. It has also made use of the Habitable Zone Gallery at hzgallery.org.

## Author Disclosure Statement

No competing financial interests exist.